\documentclass[sigconf]{acmart}

\AtBeginDocument{%
  }

\setcopyright{acmlicensed}
\copyrightyear{2025}
\acmYear{2025}
\acmDOI{XXXXXXX.XXXXXXX}
\acmConference[Conference acronym 'XX]{Make sure to enter the correct
  conference title from your rights confirmation email}{June 03--05,
  2018}{Woodstock, NY}
\acmISBN{978-1-4503-XXXX-X/2018/06}

\usepackage{caption}
\usepackage{subcaption}
\usepackage{makecell}
\usepackage{multirow}

\begin{document}

\title{Dual-disentangle Framework for Diversified Sequential Recommendation}





\author{Haoran Zhang}
\authornote{Both authors contributed equally to this research.}
\affiliation{%
  \institution{Tianjin University}
  \city{Tianjin}
  \country{China}}
\email{zhr00@tju.edu.cn}

\author{Jingtong Liu}
\authornotemark[1]
\affiliation{%
  \institution{Tianjin University}
  \city{Tianjin}
  \country{China}}
\email{liujingtong@tju.edu.cn}

\author{Jiangzhou Deng}
\affiliation{%
  \institution{Chongqing University of Posts and Telecommunications}
  \city{Chongqing}
  \country{China}}
\email{dengjz@cqupt.edu.cn}

\author{Junpeng Guo}
\authornote{Corresponding author.}
\affiliation{%
  \institution{Tianjin University}
  \city{Tianjin}
  \country{China}}
\email{guojp@tju.edu.cn}
\orcid{0000-0002-3884-3180}

\renewcommand{\shortauthors}{Liu et al.}

\begin{abstract}
Sequential recommendation predicts user preferences over time and has achieved remarkable success. However, the growing length of user interaction sequences and the complex entanglement of evolving user interests and intentions introduce significant challenges to diversity. To address these, we propose a model-agnostic \textbf{D}ual-disentangle framework for \textbf{D}iversified \textbf{S}equential \textbf{Rec}ommendation (\textbf{DDSRec}). The framework refines user interest and intention modeling by adopting disentangling perspectives in interaction modeling and representation learning, thereby balancing accuracy and diversity in sequential recommendations. Extensive experiments on multiple public datasets demonstrate the effectiveness and superiority of DDSRec in terms of accuracy and diversity for sequential recommendations. 
\end{abstract}




\maketitle

\section{Introduction}

Personalized Recommender Systems(RSs) are now widespread in areas such as online content consumption, retail, and travel, helping users discover items they might otherwise miss\cite{7927889, 9693280}. Unlike early methods that treat user behavior as isolated interactions, sequential recommendation takes advantage of the sequential nature of user behavior\cite{BOKA2024102427}. Explores the intricate evolution patterns of user interests within the sequence, thus providing more accurate and personalized recommendations\cite{aaai33015941}.

The advancement of sequence models has significantly propelled sequential recommendation systems, enabling their application in various real-world scenarios\cite{11453426723, SeqRecfcs, 1011453653016}. However, excessive focus on recommendation accuracy can trigger issues such as information cocoons, the Matthew effect, and filter bubbles\cite{10490254, 2568012, 1011453664928}, which can undermine user experience and platform revenue\cite{35893343645497, mnsc10800974, joc190952}. Industry and academia have long acknowledged the importance of temporal factors and diversity in modeling user intentions to improve the quality of recommendations\cite{7927889}. Sequential and diverse recommendation approaches address different facets of the evolution towards more personalized and intelligent RSs\cite{BOKA2024102427, KUNAVER2017154}. Sequential recommendation has become a core method in the technical framework of RSs, while diversity has emerged as a significant research topic\cite{SeqRecfcs, 1011453664928}.

Recent research has concentrated on prompting the diversity of sequential recommendations\cite{1011453653016, aaai33258}. Most relevant studies can be broadly categorized into two types: technical and strategic. Technical approaches introduce and adapt state-of-the-art sequential methods from the field of artificial intelligence to infuse diversity into the modeling phase of sequential recommendations\cite{1011453653016, cikm3557071}. Strategic approaches, on the other hand, combine existing or improved modeling methods with advanced strategies such as data augmentation\cite{aaai33258} and representation disentanglement\cite{wsdm1011453570389} to boost diversity. However, current research faces several challenges, which could be summarized as follows.

\begin{figure*}
    \centering
    \begin{subfigure}[b]{0.5\textwidth}
        \centering
        \includegraphics[width=\textwidth]{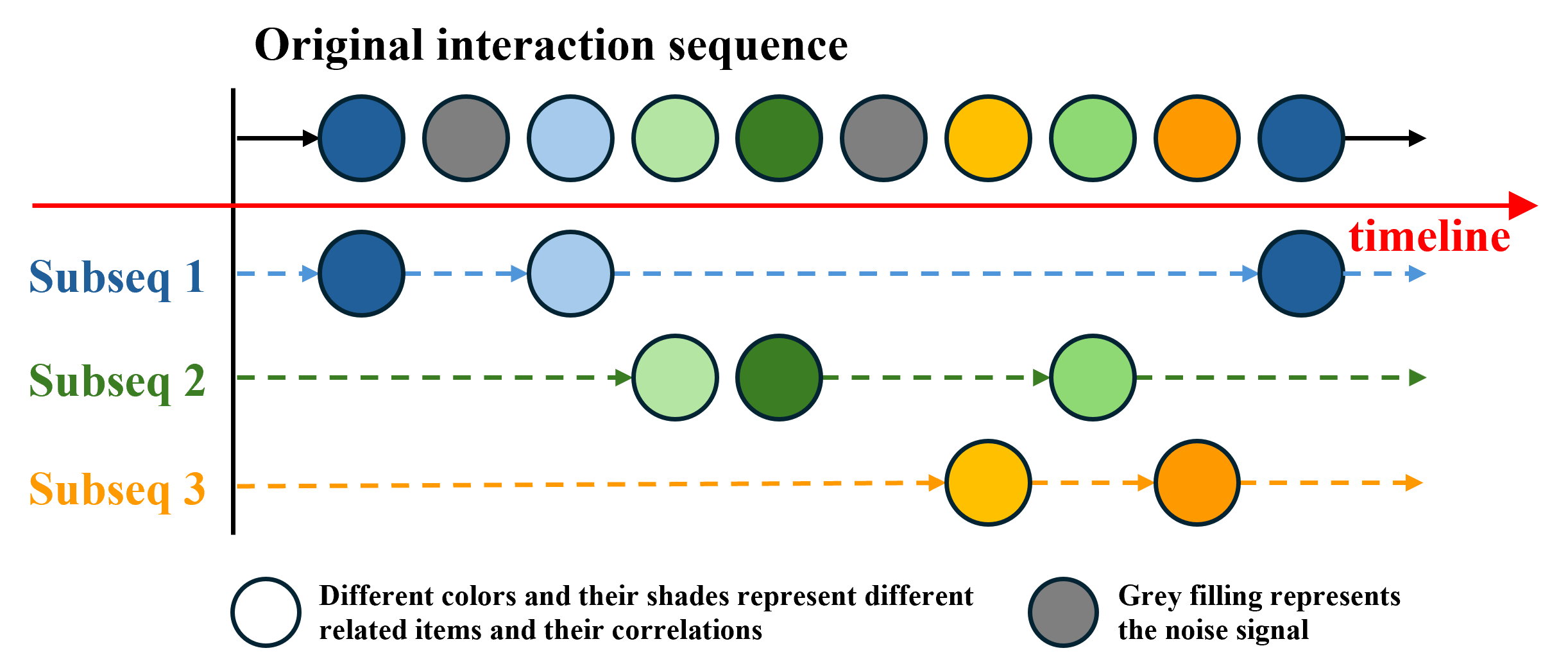}
        \caption{\footnotesize Complex interests and noise entanglement in temporal dimension.}
        \label{fig1a}
    \end{subfigure}
    ~~
    \begin{subfigure}[b]{0.4\textwidth}  
        \centering
        \includegraphics[width=0.65\textwidth]{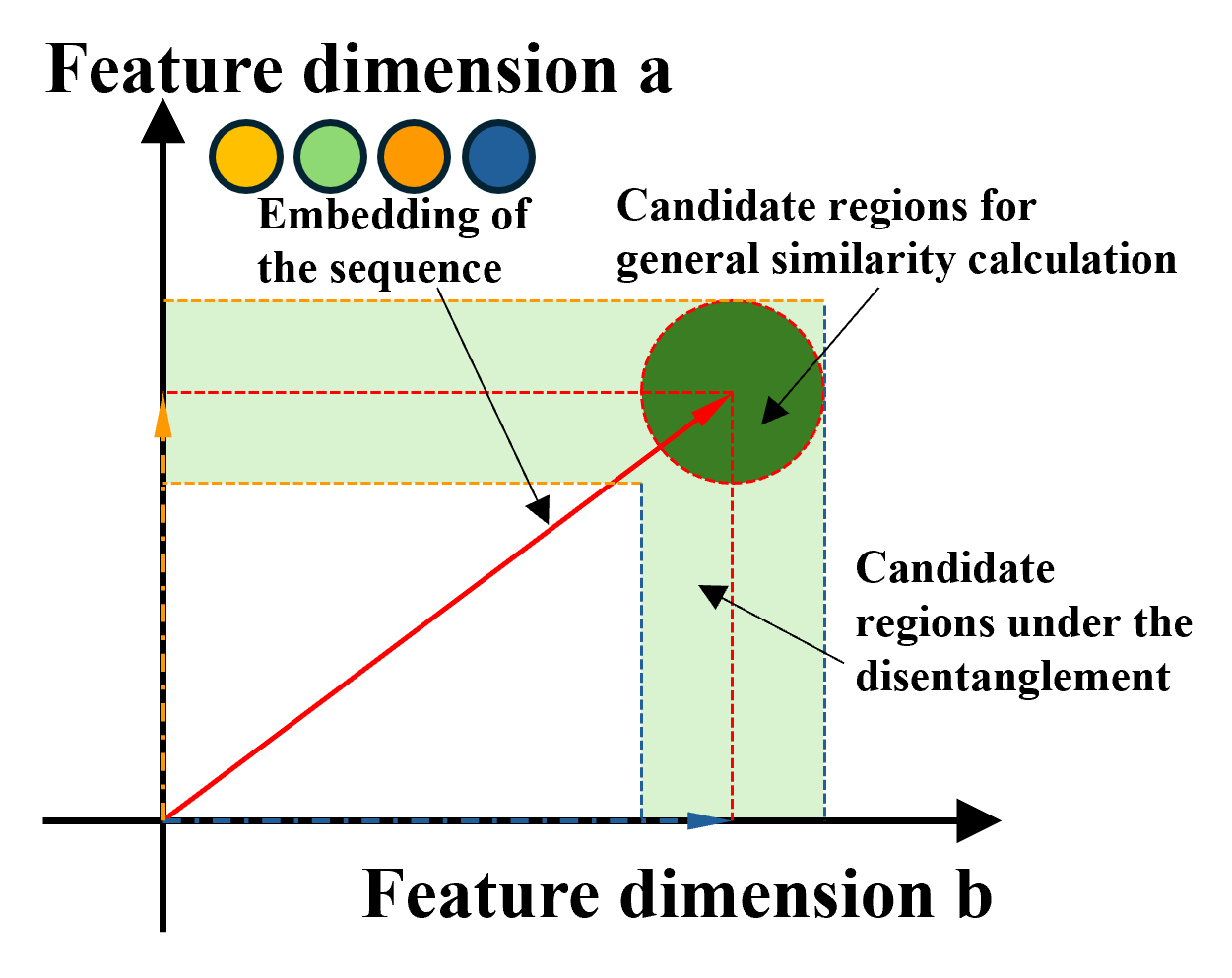}
        \caption{\footnotesize Disentangled representation captures broader user intentions.}
        \label{fig1b}
    \end{subfigure}
    
    \vspace{-0.3cm}   
    \caption{Schematic diagram of the disentangling and diversity prompt principle in two dimensions.}
    \vspace{-0.3cm}
    \label{fig1}
    \Description{figure 1}
\end{figure*}

\textbf{Entanglement of complex user interests in interaction sequence.} Although user interaction sequences unfold within a one-dimensional temporal dimension, the correlations among items in the sequence can transcend the "order" \cite{ipm2024103619}. A single user's historical interaction sequence may be viewed as a complex interplay of multiple correlated sub-interest interaction sequences and noisy interactions, as illustrated in Figure \ref{fig1a}. This characteristic is especially prominent in ultralong sequences and poses challenges for user modeling and diversity prompting in recommendations.

\textbf{Entanglement of complex user intention on item feature.} User complex intentions in historical sequences reflect the diversity of the user\cite{1011453664928}. The engagement of a user with a set of items is derived from preferences of varying strengths in different aspects of these features of the items\cite{kdd3709429}. Representation learning embeds these complex intentions into a single vector, potentially overlooking distinct user interests\cite{wsdm1011453570389}, as shown in Figure \ref{fig1b}.


\textbf{Balancing recommendation diversity and accuracy.} The relationship between recommendation diversity and accuracy is often seen as a dynamic trade-off. Some industry practice and academic research indicate an inverse relationship between them\cite{GOGNA201783, iui3712075}. Improving one can often lead to a decrease in the other, presenting a fundamental challenge in improving both diversity and accuracy in recommendation systems\cite{10891877, aaai33258}.

Our work addresses the key challenges above in sequential recommendations by making the following contributions:
\textbf{\romannumeral1.} A novel framework improves both the accuracy and diversity of sequential recommendations, providing new methodological insights.
\textbf{\romannumeral2.} Diversity is integrated in user interest modeling through disentanglement of sequence and interest representation, addressing the complexity of user interests and intentions.
\textbf{\romannumeral3.} Experiments on three public datasets demonstrate that our DDSRec outperforms advanced baselines in recommendation accuracy and diversity. Ablation studies confirm the importance of each component in our framework for effective user modeling and diversity prompt.

\section{Framework}

For user and item set $\mathcal{U}$ and $\mathcal{I}$, the sequential recommendation seek to predict the next item given a user’s($u\in \mathcal{U} $) historical interaction sequence $\mathcal{S}^u=(\mathcal{S}^u_1, \mathcal{S}^u_2, \ldots , \mathcal{S}^u_{t-1})$ at time step $t$.

\subsection{Sequence Disentanglement}

An item embedding matrix $\mathbf{M}\in\mathbb{R}^{|\mathcal{I}|\times d}$ is created first, where $d$ is the latent dimensionality. $\mathbf{M}_{i_j}$ is the embedding of the item $i_j\in\mathcal{S}^u$. Similarly to\cite{ipm2024103619}, we employ an adaptive masking module to divide the user-item interaction sequence into two parts: the trend interest sequence ($\mathcal{S}^u_m$) and the discrete interest sequence ($\mathcal{S}^u_d$). Using the most recent $p$ items as proxy $\mathbf{p}=1/p \sum_{j=t-p}^{t-1}{\mathbf{W}_s\mathbf{M}_{i_j}} $, the mask vector $\mathbf{m}=[m_1,m_2,...,m_j]$($m_j\in\{0,1\}$) is determined in the following way:
\begin{equation}
m_j=\begin{cases}
 1, if cosine\left(\mathbf{M}_{i_j},\mathbf{p}\right)\geq\theta_m\\
 0, otherwise
\end{cases}
\end{equation}
where $cosine\left(\cdot,\cdot\right)$ calculates the cosine similarity, $\mathbf{W}_s\in\mathbb{R}^{d \times d}$is learnable parameters and $\theta_m$ is a pre-defined threshold. $\mathcal{S}^u_m$ is also organized chronologically and has the same order as in $\mathcal{S}^u$.

For $\mathcal{S}^u_m$, theoretically any suitable sequential method can be used to uncover latent interest trends and their complex evolutions within the interaction sequence\cite{SeqRecfcs,1011453653016}. The transformer blocks are chosen to be adopted in our current implementation. Position embeddings $\mathbf{P}\in\mathbb{R}^{n\times d}$ are injected as:
$\widehat{\mathbf{E}}=[\mathbf{M}_{i_1}+\mathbf{P}_{1}, \mathbf{M}_{i_2}+\mathbf{P}_{2}, \dots, \mathbf{M}_{i_{n}}+\mathbf{P}_{n}]$. The multi-head attention is used to aggregate sequence representation:
\begin{equation}
\mathbf{S} = MHSA(\widehat{\mathbf{E}}) = Concat(head_1, head_2,\cdots,head_h)\mathbf{W}_H,
\end{equation}
where $head_n=SA(\widehat{\mathbf{E}})=\text{Attention}(\widehat{\mathbf{E}}\mathbf{W}_Q, \widehat{\mathbf{E}}\mathbf{W}_K, \widehat{\mathbf{E}}\mathbf{W}_V)$ 
and the scaled dot-product attention \cite{transformer} is defined as:
\begin{equation}
\text{Attention}(\mathbf{Q},\mathbf{K},\mathbf{V})=\text{softmax}\left (\frac{\mathbf{Q}\mathbf{K}^T}{\sqrt{d}}\right)\mathbf{V},
\end{equation}
where $\mathbf{Q}$ represents the queries, $\mathbf{K}$ the keys and $\mathbf{V}$ the values.
Next, point-wise Feed-Forward Networks (FFN) to further enhance the model with non-linearity:
$\mathbf{F} = FFN(\mathbf{S})$, and the $Transformer$ layer is constructed by stacking the self-building blocks and the $b$-th block is defined as:
\begin{equation}
\mathbf{S}^b = MHSA(F^{(b-1)}), \mathbf{F}^b = FFN(\mathbf{S}^b), \forall b \in {1,2,\cdots}.
\end{equation}

For $\mathcal{S}^u_d$, we utilize Multilayer Perceptron (MLP) to integrate these features, aiming to preserve the underlying interests of the users as much as possible.

\begin{figure}
    \centering
    \includegraphics[width=\linewidth]{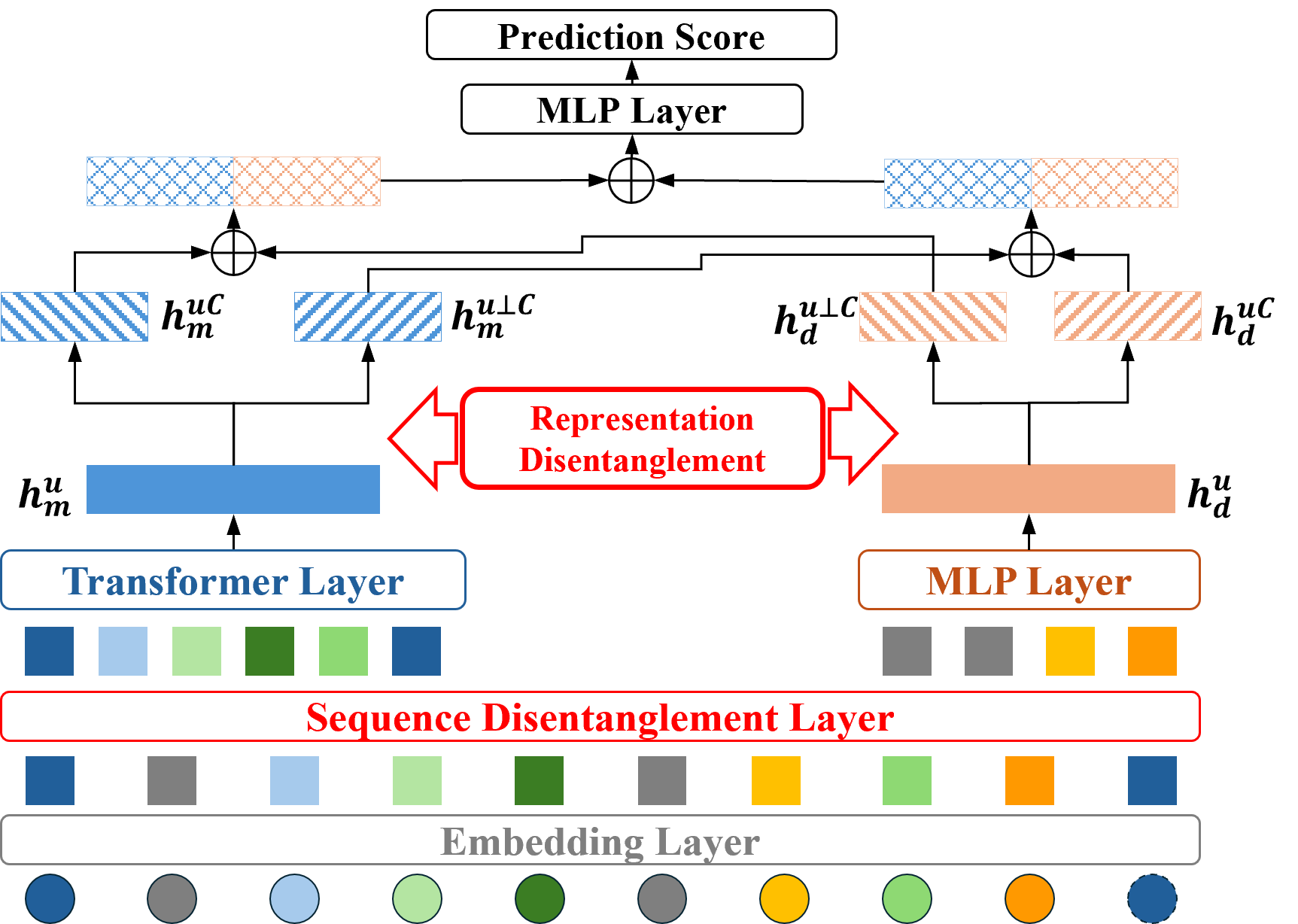}
    \caption{The overview of the Dual-disentangle Framework.}
    \label{framework}
    \Description{figure 2}
    \vspace{-0.3cm}
\end{figure}

After disentangling and encoding the interaction sequences, we derive two representation vectors: $\mathbf{h}^u_m=Transformer(\widehat{\mathbf{E}}_{[\mathcal{S}^u_m]})$, $\mathbf{h}^u_d=MLP(\mathbf{M}_{[\mathcal{S}^u_d]})$, which capture the user's interest trends and discrete interests, respectively, as depicted in Figure \ref{framework}. Subsequently, we further disentangle these two interest representations to capture different facets of the user's intent.

\subsection{Representation Disentanglement}

Previous studies have shown that user interest in a single item can be disentangled into two independent aspects: category-dependent and category-independent\cite{wsdm1011453570389}. We extend this insight to sequence representation to facilitate further diversity integration. Specifically, we construct an adversarial discriminator $D(\cdot)$, using the multi-hot encoding $\mathbf{c}_{i_t}$ of the category (or categories) of item $i_t$ as a supervisory signal. The adversarial training is enforced to meet the following objectives simultaneously:
\begin{align}
min\mathcal{L}^C_D\left(\mathcal{S}^u,i_t\right) =& \mathcal{L}_{CE}\left(D(\mathbf{h}^C), \mathbf{c}_{i_t}\right),\\
max\mathcal{L}^{\perp C}_D\left(\mathcal{S}^u,i_t\right) =& \mathcal{L}_{CE}\left(D(\mathbf{h}^{\perp C}), \mathbf{c}_{i_t}\right),
\end{align}
where $\mathcal{L}_{CE}$ is the cross entropy loss, $\mathbf{h}^C$ and $\mathbf{h}^{\perp C}$ are two independent components that encode the category related and irrelevant information inherent in the original representation, respectively. The discriminator $D(\cdot)$ is implemented with MLP. Two adversarial discriminators are used to obtain the decoupled representations of $\mathbf{h}^u_m$ and $\mathbf{h}^u_d$ respectively.

\subsection{Cross-prediction}

The four subrepresentations $\mathbf{h}_m^{uC}$, $\mathbf{h}_m^{u\bot C}$, $\mathbf{h}_d^{uC}$ and $\mathbf{h}_d^{u\bot C}$, from dual disentanglement, are merged via pairwise cross-fusion. This aims to incorporate diverse category-related interests into trend intentions while retaining useful information from discrete category-independent interests. Finally, prediction score $\widehat{\mathbf{y}}$ is generated by:
\begin{equation}
\widehat{\mathbf{y}} = 
MLP\left(
Concat\left(
MLP(\mathbf{h}_m^{uC},\mathbf{h}_d^{u\bot C}),
MLP(\mathbf{h}_m^{u\bot C},\mathbf{h}_d^{uC})
\right)\right).
\end{equation}

The training loss of the entire framework consists recommendation loss and adversarial loss:
\begin{equation}
\mathcal{L} = \mathcal{L}_{CE}\left(\widehat{\mathbf{y}},\mathbf{y}\right)
+ \lambda_1 \left(\mathcal{L}^{mC}_D + \mathcal{L}^{dC}_D \right) 
- \lambda_2 \left(\mathcal{L}^{m \bot C}_D + \mathcal{L}^{d \bot C}_D\right).
\end{equation}
where $\lambda_1$ and $\lambda_2$ are hyperparameters.

\section{Experiments and Results Analysis}

This section validates the effectiveness of our DDSRec framework through comprehensive experiments across benchmark datasets
\footnote{The code will be available from \href{https://github.com/sunreclab/cikm25}{https://github.com/sunreclab/cikm25}.}
, centering on two core research inquiries:
\textbf{RQ1}: Does the proposed method outperform the baselines in balancing recommendation accuracy with diversity?
\textbf{RQ2}: What incremental value do individual framework components contribute to overall performance?

\subsection{Datasets and Baselines}

Three publicly available datasets, \textbf{KuaiRec}\cite{cikm3557220}, \textbf{Tenrec}\cite{nips3601104}, and \textbf{MIND}\cite{mind2020}, collected from different real-world scenarios, were chosen as benchmark datasets. All datasets were subjected to 5-core filtering\cite{cikm3411897} to ensure interaction sufficiency. Historical user interactions were chronologically sorted, and the leave-one-out was applied to partition the data into train set, validation set, and test set while maintaining temporal coherence. The statistics of the three datasets after preprocessing are presented in Table\ref{tab1}.

\begin{table}[htbp]
    \setlength{\abovecaptionskip}{0.cm}
    \centering
    \caption{Statistics of the three datasets after preprocessing.}
    \label{tab1}
    \begin{tabular}{p{0.85cm}<{\centering} p{0.85cm}<{\centering} p{0.85cm}<{\centering} p{1.35cm}<{\centering} p{1.2cm}<{\centering} p{0.85cm}<{\centering}}
    \toprule
        dataset & \#users & \#items & \#interactions & \#categories & density \\
    \midrule
        KuaiRec&  1411  &3065  &216735 &31 & 5.01\%\\
        Tenrec &  24002 &2607  &339405 &42 & 0.54\%\\
        MIND   &  36505 &12975 &849440 &15 & 0.18\%\\
    \bottomrule
    \end{tabular}
    
    \vspace{-0.3cm}
\end{table}

DDSRec was compared to five baselines that include sequential recommendation methods : \textbf{SASRec}\cite{icdm8594844}, \textbf{DIEN}\cite{aaai33015941}, \textbf{TEDDY}\cite{ipm2024103619},and diversity-enhancing approaches:  \textbf{DGRec}\cite{wsdm3570472} and \textbf{DCRS}\cite{wsdm1011453570389}. These baselines encompass state-of-the-art methodologies such as attention, graph, and disentangle methods.

\subsection{Implementation and Evaluation}

For all three datasets, we set the dropout rate to 0.1, the embedding dropout rate to 0.3, and the embedding size to 64, using Adam as optimizer. The loss weights for representation disentanglement are (0.5, 0.5). For fair comparison, we used official implementations of all baseline models and tuned them using the validation set. Where applicable, we retained all category-aware components in baseline methods to align with our model's use of item category information to enhance recommendation diversity.

The metrics for recommendation accuracy (Recall@\textit{K}, NDCG@\textit{K}) and diversity (CE@\textit{K}, CC@\textit{K}) were used for the evaluation, with K in \{5, 10, 20\}. Recall@\textit{K} measures the proportion of correctly predicted next-click items in top-\textit{K} recommendations, while NDCG@\textit{K} assesses their ranking quality. Higher values in both indicate better accuracy. CE@\textit{K} reflects the entropy of the category distribution among top-\textit{K} items, and CC@\textit{K} shows the proportion of unique categories covered. Higher values signify better diversity. 



\subsection{Results and Analysis}

\subsubsection{Overall Comparison (\textbf{RQ1})}

    

The results of the comparative experiment are shown in Table \ref{results}. In most cases, DDSRec demonstrates significantly enhanced recommendation accuracy and diversity compared to baseline models. 

In the few cases where it does not exceed certain models, it still achieves a balanced performance. For instance, on the KuaiRec dataset, Recall@5 of DDSRec is second only to SASRec, yet it excels in all other diversity metrics. The reason is that they used similar methods in sequence encoding, but DDSRec has a series of diversity fusion operations. Although DCRS slightly surpasses DDSRec in the CE@20 metric, this comes at the cost of significant accuracy loss. Other situations are similar to these two examples. 

In general, DDSRec enhances both recommendation accuracy and diversity, effectively addressing the balancing challenge between them.

\begin{table*}[!ht]
    \setlength{\abovecaptionskip}{0.cm}
    \centering
    \caption{Comparative experiment results. }
    \label{results}
    \footnotesize

    \begin{tabular}{p{0.9cm}<{\centering} p{0.9cm}<{\centering} p{0.85cm}<{\centering} p{0.85cm}<{\centering} p{0.85cm}<{\centering} p{0.85cm}<{\centering} p{0.85cm}<{\centering} p{0.85cm}<{\centering} p{0.85cm}<{\centering} p{0.85cm}<{\centering} p{0.85cm}<{\centering} p{0.85cm}<{\centering} p{0.85cm}<{\centering} p{0.85cm}<{\centering}}
        \hline 
        dataset & method & Recall@5 & Recall@10 & Recall@20 &  NDCG@5  &  NDCG@10 &  NDCG@20 &  CE@5  &  CE@10  &  CE@20  &  CC@5  &  CC@10  &  CC@20  \\
        \hline 
          \multirow{7}{*}{KuaiRec} & SASRec &\textbf{12.3883} &15.2870 &18.1526 &\underline{7.9610 }&\underline{8.8922} &9.6785  &1.2708 &1.6975 &2.0363 &0.1302 &0.2184 &0.3234\\
           & DIEN   &11.0415 &14.6877 &17.9313 &7.5372 &8.8849 &9.2510 &\underline{1.2963} &1.7084 &2.0212 &\underline{0.1374} &\underline{0.2201} &0.3179\\
           & DGRec  &4.9618 &8.2601 &13.8200 &2.6650 &4.2623 &5.8706 &1.2905 &\underline{1.7911} &2.1397 &0.1339 &0.2087 &0.3354\\
           & DCRS   &4.1000 &7.5800 &13.1800 &2.4300 &3.5400 &4.9600 &1.2010 &1.6879 &\textbf{2.3133} &0.1163 &0.1923 &\underline{0.3438}\\
           & TEDDY  &9.9246 &\underline{15.2948} &\textbf{21.8909} &7.0444 &8.8014 &\underline{10.4605} &1.2418 &1.7128 &2.1334 &0.1299 &0.2135 &0.3296 \\
           & \textbf{DDSRec} &\underline{11.2857} &\textbf{16.1962} &\underline{21.6152} & \textbf{8.0147}&\textbf{9.6047} &\textbf{10.9936} &\textbf{1.3678} &\textbf{1.8379} &\underline{2.2219} &\textbf{0.1403} &\textbf{0.2283} &\textbf{0.3451} \\
           \cline{2-14}
           & \textit{Improv.} & - &5.89\% & - &6.75\%  &8.01\%  &5.10\%  &5.52\%  &2.61\%  & - &2.11\%  &3.73\%  &0.38\%  \\       
        \hline 
     \end{tabular}

    \begin{tabular}{p{0.9cm}<{\centering} p{0.9cm}<{\centering} p{0.85cm}<{\centering} p{0.85cm}<{\centering} p{0.85cm}<{\centering} p{0.85cm}<{\centering} p{0.85cm}<{\centering} p{0.85cm}<{\centering} p{0.85cm}<{\centering} p{0.85cm}<{\centering} p{0.85cm}<{\centering} p{0.85cm}<{\centering} p{0.85cm}<{\centering} p{0.85cm}<{\centering}}
          \multirow{7}{*}{Tenrec} & SASRec &13.4080 &17.3150 &22.5347 &8.7801 &10.0301 &11.4311 &1.0584 &1.4803 &1.7931 &0.0806 &0.1318 &\underline{0.2001} \\
           & DIEN   &12.5847 &16.7930 &22.0718 &8.6024 &9.9952 &11.4219 &1.0832 &\underline{1.4891} &1.7926 &0.0837 &0.1345 &0.1994 \\
           & DGRec  &8.1660 &11.9907 &17.5069 &5.5311 &6.7645 &8.1443 &0.9700 &1.4536 &\underline{1.7935} &0.0721 &\underline{0.1351} &0.1942 \\
           & DCRS   &6.4000 &10.5100 &16.7600 &3.7500 &5.0700 &6.6300 &\textbf{1.1334} &1.4872 &1.7368 &\underline{0.0842} &0.1296 &0.1882 \\
           & TEDDY  &\underline{15.4628} &\underline{23.0213} &\underline{34.3146} &\textbf{10.7470} &\underline{13.1764} &\textbf{16.0201} &0.9238 &1.4128 &1.7658 &0.0724 &0.1271 &0.1982 \\
           & \textbf{DDSRec} &\textbf{15.6193} &\textbf{23.2862} &\textbf{34.4629} &\underline{10.7259} &\textbf{13.1894} &\underline{15.9965} &\underline{1.1246} &\textbf{1.5612} &\textbf{1.8340} &\textbf{0.0846} &\textbf{0.1399} &\textbf{0.2065 }\\
           \cline{2-14}
           & \textit{Improv.} &1.01\%  &1.15\%  &0.43\%  & - &0.10\%  & - & - &4.84\%  &2.26\%  &0.48\%  &3.55\%  &3.20\%  \\
        \hline
     \end{tabular}

    \begin{tabular}{p{0.9cm}<{\centering} p{0.9cm}<{\centering} p{0.85cm}<{\centering} p{0.85cm}<{\centering} p{0.85cm}<{\centering} p{0.85cm}<{\centering} p{0.85cm}<{\centering} p{0.85cm}<{\centering} p{0.85cm}<{\centering} p{0.85cm}<{\centering} p{0.85cm}<{\centering} p{0.85cm}<{\centering} p{0.85cm}<{\centering} p{0.85cm}<{\centering}}
          \multirow{7}{*}{MIND} & SASRec &9.1120 &12.0200 &15.5440 &5.7474 &6.6814 &7.6343 &1.0986 &1.4273 &1.6958 &0.2281 &0.3518 &0.5156 \\
           & DIEN   &7.9633 &10.8397 &14.1785 &5.4392 &6.3790 &7.5016 &1.1139 &1.4458 &1.7147 &\underline{0.2290} &0.3536 &0.5174 \\
           & DGRec  &3.2272 &7.7861 &12.8791 &1.7502 &3.5295 &5.2016 &\underline{1.1437} &1.4201 &1.6426 &0.2279 &0.3416 &0.4803 \\
           & DCRS   &2.0700 &3.4500 &9.6200 &1.3500 &2.6000 &3.8100 &0.9529 &1.2574 &1.4964 &0.2006 &0.2986 &0.4200 \\
           & TEDDY  &\underline{9.5681} &\underline{14.9524} &\textbf{22.0976} &\underline{6.2471} &\underline{7.9738} &\textbf{9.7710} &1.1014 &\underline{1.4507} &\underline{1.7303} &0.2276 &\underline{0.3550} &\underline{0.5250} \\
           & \textbf{DDSRec} &\textbf{9.6011} &\textbf{15.0281} &\underline{21.8538} &\textbf{6.2610} &\textbf{8.0062} &\underline{9.7278} &\textbf{1.1585} &\textbf{1.4978} &\textbf{1.7653} &\textbf{0.2380} &\textbf{0.3674} &\textbf{0.5360} \\
           \cline{2-14}
           & \textit{Improv.} &0.34\%  &0.51\%  & - &0.22\%  &0.41\%  & - &1.29\%  &3.25\%  &2.02\%  &3.93\%  &3.49\%  &2.10\% \\
        \hline
     \end{tabular}
     
     \begin{tabular}{c}
        \textit{Improv.} indicates the percentage improvement over the second-best performance, if applicable; the results of Recall@\textit{K} and NDCG@\textit{K} are percentages. 
    \end{tabular}
  \vspace{-0.3cm}
\end{table*}

\subsubsection{Ablation Study (\textbf{RQ2})}

An ablation study was performed on the KuaiRec by removing various modules from the proposed framework. Specifically, the following model variants were evaluated: \textbf{\romannumeral1. \textit{w/o DD}}: Both the sequence disentanglement and intention representation disentanglement operations were removed. The entire interaction sequence was directly modeled and the learned intention representation was used to predict the next interaction. \textbf{\romannumeral2. \textit{w/o SD}}: The sequence disentanglement operation was removed, and the entire interaction sequence was directly modeled with sequential methods. Retain the representation disentanglement module. \textbf{\romannumeral3. \textit{w/o RD}}: representation disentanglement was removed and the two intention representations obtained from the sequence disentanglement module were concatenated directly for recommendation. Table \ref{Ablation study} presents the results of the ablation studies.

Removing both key operations from DDSRec simultaneously led to a significant drop in accuracy and a deterioration in diversity metrics.
Eliminating the subsequence disentanglement operation alone decreased both accuracy and diversity, underscoring the importance of subsequence decomposition. In contrast, removing the intention representation disentanglement operation resulted in slightly higher Recall@10, Recall@20, and NDCG@20 but lower diversity metrics compared to DDSRec. This indicates that while disentangling intention representations effectively enhances diversity, it may sometimes cause minor reductions in accuracy. These results confirm that both operations are crucial to achieving a balanced improvement in recommendation performance.





\begin{table}[htbp]
    \setlength{\abovecaptionskip}{0.cm}
    \centering
    \caption{Ablation study on KuaiRec.}
    \label{Ablation study}
    \footnotesize
    \begin{tabular}{p{0.8cm}<{\centering} p{0.85cm}<{\centering} p{0.85cm}<{\centering} p{0.85cm}<{\centering} p{0.85cm}<{\centering} p{0.85cm}<{\centering} p{0.85cm}<{\centering}}
        \toprule
          & Recall@5 & Recall@10 & Recall@20 &  NDCG@5  &  NDCG@10 &  NDCG@20 \\
        \midrule
          \textit{w/o. DD} &9.9419 &14.9846 &21.4056 &7.0461 &8.6783 &10.3012  \\
          \textit{w/o. SD} &10.5075 &15.5644 &\underline{22.5023} &7.2679 &8.9087 &10.6572 \\
          \textit{w/o. RD} &\underline{11.1077} &\textbf{16.4546} &\textbf{22.5257} &\underline{7.8346} &\underline{9.5325} &\textbf{11.0391} \\
          DDSRec &\textbf{11.2857} &\underline{16.1962} &21.6152 &\textbf{8.0147} &\textbf{9.6047} &\underline{10.9936} \\
        \bottomrule
    \end{tabular}
    \begin{tabular}{p{0.8cm}<{\centering} p{0.85cm}<{\centering} p{0.85cm}<{\centering} p{0.85cm}<{\centering} p{0.85cm}<{\centering} p{0.85cm}<{\centering} p{0.85cm}<{\centering}}
        \toprule
          & CE@5  &  CE@10  &  CE@20  &  CC@5  &  CC@10  &  CC@20  \\
        \midrule
          \textit{w/o. DD} &\underline{1.2803} &\underline{1.7756} &\underline{2.1833} &\underline{0.1334} &\underline{0.2208} &\underline{0.3355} \\
          \textit{w/o. SD} &1.2054 &1.6786 &2.0782 &0.1279 &0.2102 &0.3248 \\
          \textit{w/o. RD} &1.2635 &1.7207 &2.1295 &0.1314 &0.2134 &0.3296 \\
          DDSRec &\textbf{1.3678} &\textbf{1.8379} &\textbf{2.2219} &\textbf{0.1403} &\textbf{0.2283} &\textbf{0.3451} \\
        \bottomrule
     \end{tabular}

    
  \vspace{-0.3cm}
\end{table}

\section{Related Work}


\subsection{Sequential Recommendation}

Initially, frequent pattern mining and Markov models were used to model recommendation sequences\cite{11453426723}. Subsequently, with the remarkable success of sequential models like RNN, CNN, GRU, attention mechanisms, GNN, GAN and transformers in other domains, these models were also successfully integrated into recommender systems. Their integration enables a more refined and efficient modeling of user interests and intentions\cite{BOKA2024102427, 10891877}. As user-item interaction sequences become longer, retrieval-augmented and memory-augmented methods have emerged to adapt to the evolution of user behavior, interactions, and intentions in ultralong sequences\cite{SeqRecfcs, cikm3557071}.

Although the literature on sequential recommendation is extensive, existing methods still face numerous challenges. Conventional sequence methods struggle to reconcile long-term and short-term user engagement. Typical deep sequential approaches are incapable of disentangling intricate user interests. Moreover, the non-mainstream yet valuable discrete user interests hidden in "noise" present diversity challenges to recommendation sequence denoising methods based on data augmentation and retrieval. Beyond technical issues, application-level problems, such as diversity\cite{ijcai2019p380, cikm3411897, cikm3557071, 1011453653016, aaai33258}, are gaining increasing attention.

\subsection{Diversified Recommendation}

Research on recommendation diversity dates back to the collaborative filtering era\cite{KUNAVER2017154}. During the past 20 years, researchers have systematized the concept of diversity from various perspectives. These include individual and aggregate diversity\cite{joc190952}, the implementation of diversity at different stages of recommender systems (such as pre-processing, modeling, and re-ranking)\cite{cikm3614848}, as well as user diversity and item diversity\cite{1011453664928}. Some studies focus on uncovering the principles or patterns of recommendation phenomena, such as filter bubbles, from a phenomenological perspective\cite{2568012, 1011453664928}. In addition, distinctions and connections between the diversity of recommendations and related concepts, such as fairness of recommendations, have been analyzed\cite{1011453664928}. These insights provide a theoretical foundation for the design of recommendation algorithms\cite{YU2019113073, 10490254}.

A major challenge in implementing diversity in recommender systems is balancing accuracy and diversity\cite{GOGNA201783}. Although prior studies have addressed this issue\cite{UAMP3394858, ZANON2022109333}, it persists in both general and sequential recommendations\cite{iui3712075, aaai33258}.

\section{Conclusion}

In this paper, we propose DDSRec, a model agnostic dual disentanglement framework that balances accuracy and diversity in sequential recommendation. DDSRec first disentangles user interaction sequences into trend and discrete interest sequences. Then it decomposes the interest representations within each sequence into orthogonal dimensions. Finally, it predicts user preferences using cross-disentangled representations generated by these dual operations. 
Experiments on three real-world datasets show that DDSRec increases recommendation diversity while preserving accuracy. Further analysis confirms the effectiveness of both disentanglement methods. Future work will explore new generative sequence disentanglement approaches and conduct online experiments.



\begin{acks}
This study is supported by the National Natural Science Foundation of China (No.72171165).
\end{acks}

\section*{GenAI Usage Disclosure}

No GenAI tools were used.

\bibliographystyle{ACM-Reference-Format}
\bibliography{cikm}

\end{document}